\documentclass[conference]{IEEEtran}
\IEEEoverridecommandlockouts
\usepackage{cite}
\usepackage{amsmath,amssymb,amsfonts}
\usepackage{algorithmic}
\usepackage{graphicx}
\usepackage{textcomp}
\usepackage{xcolor}
\usepackage{booktabs}

\def\BibTeX{{\rm B\kern-.05em{\sc i\kern-.025em b}\kern-.08em
    T\kern-.1667em\lower.7ex\hbox{E}\kern-.125emX}}
\begin{document}

\title{MDCNN-SID: Multi-scale Dilated Convolution Network for Singer Identification\\
\thanks{$\ast$Corresponding author: Jianzong Wang (jzwang@188.com).}
}

\author{\IEEEauthorblockN{Xulong Zhang, Jianzong Wang$^{\ast}$, Ning Cheng, Jing Xiao}
\IEEEauthorblockA{\textit{Ping An Technology (Shenzhen) Co., Ltd., China} }
}
\maketitle

\begin{abstract}
Most singer identification methods are processed in the frequency domain, which potentially leads to information loss during the spectral transformation. In this paper, instead of the frequency domain, we propose an end-to-end architecture that addresses this problem in the waveform domain. An encoder based on Multi-scale Dilated Convolution Neural Networks (MDCNN) was introduced to generate wave embedding from the raw audio signal. Specifically, dilated convolution layers are used in the proposed method to enlarge the receptive field, aiming to extract song-level features. Furthermore, skip connection in the backbone network integrates the multi-resolution acoustic features learned by the stack of convolution layers. Then, the obtained wave embedding is passed into the following networks for singer identification. In experiments, the proposed method achieves comparable performance on the benchmark dataset of Artist20, which significantly improves related works.
\end{abstract}


\section{Introduction}

With the explosively increasing number of music data, it is hard for us to retrieve the desired song quickly. Therefore, the content-based automatic analysis of the song becomes essential for music information retrieval (MIR). Singer identification is considered an increasingly important research topic in MIR, the goal of which is to recognize who sang a given piece of a song. The research of singer identification is firstly proposed to apply to the music library management~\cite{sharma2019importance, csmt2021sun,panda2020multi, kooshan2019singer,zhang2021singer}. The trained singer identification model can also be used in downstream singing-related applications, such as similarity search, playlist generation, or song synthesis \cite{lee2019learning,tang2022avqvc,liu2019score,wang2022drvc,humphrey2018introduction,gao2021vocal,zhao2022nnspeech}.

The singer as an artist is critical meta-information of the song plays an important role in discriminating a song. Almost all karaoke systems and music stores categorize their music databases by using the name of artists. Additionally, singer identification can also be used for song recommendations based on similar singers and managing unlabeled songs with digital rights for massive music resources~\cite{zhang2003automatic,zhang2022Singer}. Although most of the audio files in the standard music collection contain artist tag information within metadata, audio files and metadata cannot be directly obtained in several situations, such as extracted music clips from films or television shows and recordings from live concerts. Hence, automatic singer identification is a significant and valuable task in MIR. 

However, two main factors make this task very challenging. One is that the number of music artists is enormous, and the number of songs performed by each artist is unbalanced in the real-world music datasets. The other is that the singing voice to be recognized is inevitably intertwined with the accompaniment which makes the classification more difficult~\cite{hsieh2020addressing,zhang2022MetaSID}.


During the last two decades, the research in music artist classification can be categorized into two classes \cite{ellis2007classifying}. The first kind of methods \cite{2011Charbuillet,2019Zhang_SID_UBM,aolan2021, 2019Lee-Representation} are based on traditional machine learning, which mainly focuses on feature engineering. The raw audio data are divided into short frames in which acoustic features are calculated in the frequency domain as the input data of the model. The input data is used to train a classifier such as the Gaussian Mixture Model (GMM) \cite{2007Ellis}, KNN \cite{2015Ratanpara}, MLP \cite{2014Hu} and so on. Ellis~\textit{et al.}~\cite{2007Ellis} investigated beat-synchronous and instrumental-invariant Chroma feature, which is designed to reflect melodic and harmonic content. Finally, the frame-level features MFCC and Chroma are combined with Gaussian as for the classifier, the combined feature's accuracy is 0.57 which has a lot of space to improve. 

While in the methods mentioned above, the context information of the music in frame-level features is ignored. To solve the problem, song-level audio features extracted from the whole clip are introduced to singer identification. In~\cite{2015ivector}, Eghbal-zadeh~\textit{et al.} proposed a song-level descriptor, i-vectors, which is calculated by using the frame-level timbre features MFCCs. The i-vectors provide a low-dimensional and fixed-length representation for each song and can be used in a supervised and unsupervised manner.

The other approach is deep neural network-based methods without a complicated design of handcraft features. In the research of Snyder~\textit{et al.}~\cite{snyder2018x}, they proposed a novel method to keep a comparable performance in the task of speaker recognition called x-vector, which embedded fixed-dimensional DNN. In 2019, Nasrullah~\textit{et al.}~\cite{2019CRNN} tried to use Mel-Frequency Cepstrum Coefficients (MFCCs) as the input, and a stacked Convolutional and Recurrent Neural Network (CRNN) model is built to learn the mapping between MFCCs and their corresponding artist. Although the model can learn the context information from the time-frequency features, the transformation from raw audio data to a matrix with just 13 dimension coefficients of MFCC, unavoidably leads to information loss. 

Previous works for singer identification mostly take time-frequency representation as input, which can better balance the raw information and feature size. The process of short-time Fourier transform (STFT) omits some parts of the signal by default, \textit{i.e.}, phase information. Additionally, the STFT output depends on several parameters such as the length and the overlap of the frames \cite{griffin1984signal,zhang2022TDASS}, which is fixed in the transformation process and may not be the best choice for music source separation \cite{lluis2018end}.

\begin{figure*}[ht]
	\centering
	\includegraphics[width=0.65\textwidth]{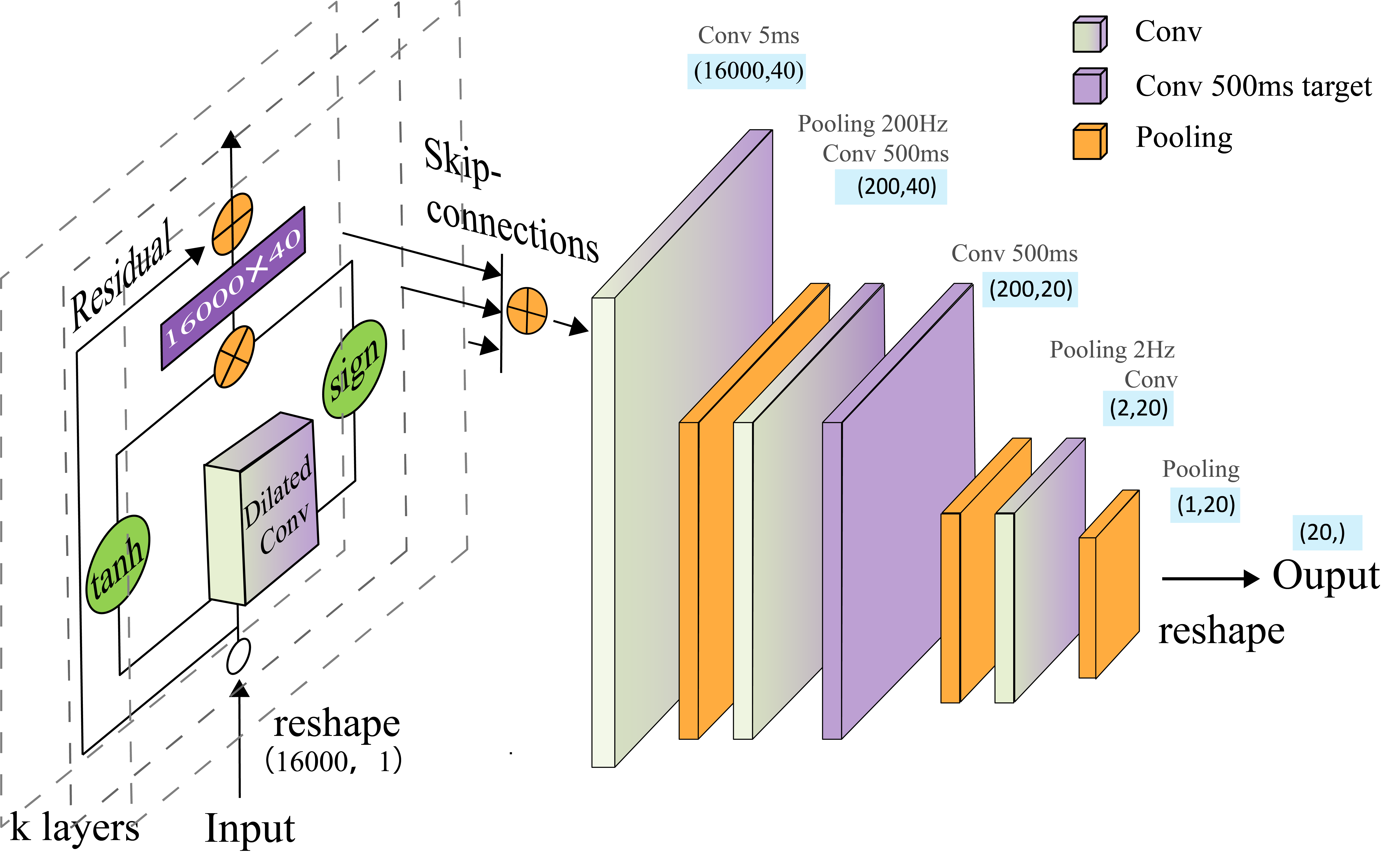}
	\caption{Overview of the MDCNN-based deep model architecture. The left part from input to skip-connections is multi-scale dilated convolution layers with several stacked layers, with the purpose of extracting high-level acoustic features. And the right part from skip-connection to output is feed forward network used for artist classification.}
	\label{fig:Model Architecture}
\end{figure*}


Motivated by the WaveNet used as a classifier in voice activity detection \cite{2019vad}. In this paper, we propose an end-to-end artist classification method in the time-domain, in which the feature extraction block is directly based on raw audio waveform. To solve the problem that frame-level acoustic features can not reflect the context information of music, we use a multi-scale dilated convolution neural network~(MDCNN) to extract high-level features of songs. Compared with ordinary feed-forward networks or Convolutional Neural Networks (CNN)~\cite{lecun1998gradient,zhang2022SUSing}, MDCNN is better in handling the long-term temporal dependencies that exist in audio signal~\cite{2019vad}, which is also quite different from another song-level descriptor i-vector~\cite{2015ivector} that is based on the spectral signal. To evaluate the effectiveness of our methods, we conduct the experiments on artist20 and our self-made dataset Singer107. Experiments show that the proposed method achieves superior results compared to the baseline algorithms~\cite{snyder2018x,2015ivector}.

Reiterating, the contributions of the paper are,
\begin{itemize}
\item We propose an end-to-end architecture for singer identification in the waveform domain, which ensures our system can learn more complete features.

\item An encoder based on MDCNN was introduced to generate wave embeddings for the raw audio signal.

\end{itemize}

\section{Related Works}
Raw waveform acoustic modeling was attracted more and more researchers in speech and music-related processing tasks over the recent several years~\cite{loweimi2020robustness,fu2017raw,qubo2021}. Sainath \textit{et al.}~\cite{sainath2015learning} proposed to use Convolutional, Long Short-Term Memory Deep Neural Network~(CLDNN) trained with raw waveform features over 2000 hours of speech could comparable with the performance of log-mel filterbank energies. The CLDNN is the time convolution layer in reducing temporal variations, through the frequency convolution preserve the locality and reduce the frequency variable. And the addition of the LSTM layer can model the temporal relation for the raw waveform data. The model directly trained on raw waveform can be tread same as the function of learning filter banks. Zeghidour \textit{et al.}~\cite{zeghidour2018learning} train a bank of complex filters that operates on the raw waveform data and then feed it into a convolution neural network (CNN). The time-domain filterbanks were processed as the same as the mel filterbanks, and then jointly finetuned with the CNN. The experiment of the phone recognition task shows the performance can outperform the CNN model directly on mel filterbanks. Ravanelli~\textit{et al.}~\cite{ravanelli2018speaker} proposed SincNet, which is a novel CNN. SincNet is based on the parametrized sinc function which acts as the band-pass filter. From the raw waveform data directly lean the low and high cutoff frequencies. The experiment on speaker identification show it was better than the standard CNN.

Different from most audio process tasks, raw waveform data does not need preprocessing to get handcraft features such as MFCC and PLP. Ghahremani \textit{et al.}~\cite{ghahremani2016acoustic} involves the feature extractor as a part of the network to do joint training. With a convolution layer to operate on a short clip of audio data with a step size of about 1.25 milliseconds, then aggregate the filter output over a fixed duration of the time axis, finally do a downsample with the rest network. The experiment shows that direct from the signal is competitive with the network based on traditional features with ivector adaptation. With the stride convolution layer, the model is expected to learn a filter bank over the raw waveform data such as frame-level features. It is because the stride size, filter size, and the number of the filters in the first convolutional layer corresponds to the hop size, windows size, and the number of melbands in the mel spectrum. In the work of Kim \textit{et al.}~\cite{kim2018sample}, SampleCNN was proposed to use 1D convolutional as the prenet of other network architecture such as ResNets and SENets. The experiment results show it could achieve state-of-the-art on the auto-tagging task. The main network is based on CNN for the raw waveform input. The majority of them used large-sized filters in the first convolutional layer, and various stride sizes to capture the frequency response.

\section{Method}

\subsection{Architecture}
The proposed network architecture is shown in Figure \ref{fig:Model Architecture}, which contains a feature encoder and an artist classification block. The proposed model takes the fixed-length segments(on this paper, the length is 16000) as input and feeds the input into the MDCNN layers. Given the waveform embedding obtained from the above feature extraction block, the convolution neural network is built to discriminate which artist the input belongs to. The detail of our model will be discussed below.

The encoder MDCNN consists of several stacked residual blocks with dilated convolutions, which makes the encoder have a sizeable receptive field without increasing the computational cost. The dilated convolution takes the filter used in a region more significant than its original size by skipping input values with a predetermined dilation factor. Skipping the input values enables the inflated convolution to have a larger receptive field than the standard convolution. Several dilated convolutions are then stacked by skip connection to integrate multi-resolution acoustic features further.


Once the MDCNN layers obtain the audio signal's embedding, the representation is then fed into the classification layers. Classification layers are built by stacking several convolution layers and pooling layers, as shown on the right part of Figure \ref{fig:Model Architecture}. All three pooling layers are used to do the downsampling operation on the raw waveform data. For example, The input with the shape of (16000,40) is changed into the hidden code of (200,40) by the first pooling layer while the second pooling layer changed the hidden code with the shape of (200,20) to a new hidden code of (2,20). between the two pooling layer is the common convolution layers.  


\begin{figure}[t]
	\centering
	\includegraphics[width=0.47\textwidth]{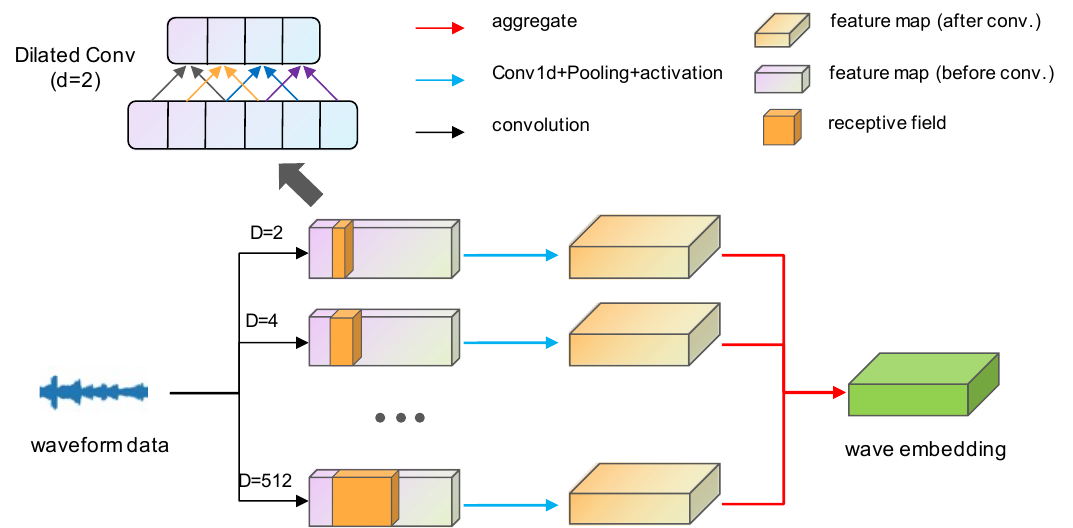}
	\caption{The illustration of Multi-scale dilated convolution. The waveform data is fed into the several parallel dilated convolution layers with different dilated factors (exponentially increasing in the proposed method). After activation, the multiple feature maps are then aggregated by skip connection to obtain the wave embedding.}
	\label{fig:multi-scale conv}
\end{figure}

\subsection{Multi-scale Dilated Convolution}

To further increase the receptive field of convolution layers and integrate acoustic features extracted by multiple resolutions, we introduce Multi-scale dilated convolution networks (MDCNN) as the feature encoder. MDCNN is the backbone of WaveNet \cite{2016wavenet}, which is firstly proposed for Text to Speech (TTS) system with timbre features along to speaker. The WaveNet is an auto-regressive network that directly estimates a raw waveform from the sample point in the temporal domain.

In \cite{2015rawwave}, the WaveNet architecture is proposed for speech music generation in the time domain, where the predicted distribution of each audio sample is conditioned on its previous audio samples.

In \cite{2016wavenet}, the vital part of the WaveNet, \textit{i.e.}, Multi-scale Dilated Convolution (MDC) layers were used to extract high-level features from the input wave while a stack of many pooling and convolution layers was used to establish a discriminant model which achieves considerable results for the phoneme recognition task. In recent research, the MDC is popularly used in speech tasks such as voice activity detection \cite{2019vad} and speech augmentation \cite{2019augment,sibo2022}.

 As shown in Figure \ref{fig:Model Architecture}, Given the waveform sequence $x = x_1,..., x_N$ as input, the model estimates the joint probability of the signal as follows:

\begin{equation}
	p(x)=\prod_{n=1}^Np(x_n|x_{n-R},x_{n-R+1},\dots,x_{n-1},\Delta)
\end{equation}

\noindent where $\Delta$ represents model parameters and $R$ represents the receptive field length. The architecture of MDC includes gated activation and dilated convolution. Within a residual block, the gated activation function is defined as:

\begin{equation}
	z=\tanh({W_{f,l}\ast x})\odot\delta(W_{g,l}\ast x)
\end{equation}

\noindent where $\odot$ represents the element-wise product operator, $\ast$ is a causal convolution operator, $l$ denotes the layer index, $g$ and $f$ represent a gate and a filter, respectively, and $W$ denotes a trainable convolution filter. The number of residual blocks and the configuration of convolution kernel in the MDC are adapted directly from prior work \cite{oord2017parallel}.

Here, We use $V$ to represent the output wave embeddings and let k represent the number of parallel convolution layers. The skip connections are performed from the output of all residual blocks as shown like $V = \sum_{i=1}^{k}z_i$.

\subsection{Model Configuration}

The waveform data of each track is divided into fixed-length blocks in the time domain as the input of the model. The output of the model is configured according to the number of singers in the dataset. According to the ground truth of each track, all segments are labeled.

The MDCNN in this paper contains a residual block, which is constructed by stacking 9 dilated convolution layers with a certain exponentially increasing dilated factor (from 2 to 512 in this block). A one-dimensional filter with a size of 2 is used for all regular convolutions and dilated convolutions throughout the network. We empirically set the channel length of a dilated convolutional layer with a fixed size of 40. The outputs of all residual blocks are then added up and fed into a one-dimensional regular convolution with a filter number of 16000 and kernel size of 40. Therefore, the dimension of the obtained feature map is (16000, 40). Subsequently, an adaptive one-dimensional average pooling layer is finally used which operates in the time domain to further aggregate the activation output of all residual blocks. And the ReLU function is adopted as an activation function for each convolution layer.

The input size and the number of filters of convolution operations in residual blocks are fixed by the experiment with grid search in a range. After that, the output wave embeddings are extracted to be passed into the CNN for classification. The number of convolutional filters and layers in the CNN classifier is modified from the setting in the prior work \cite{2019Lee-SampleCNN}. The output size is the same as the number of the artist for classification.



\section{Experiments}

\subsection{Dataset}
Totally two datasets are used in the experiments for evaluation, including one public dataset (Artist20) and one self-made dataset (Singer107).

The Singer107 dataset was collected from the online music service, which contains a total of 3262 tracks and 107 singers with 3 albums each. The dataset Artist20 \cite{2007Ellis} consists of 1413 MP3 tracks from 20 artists. Two datasets are both split by song level, which is according to the ratio of 8:1:1 into a training set, validation set, and test set. 

\subsection{Evaluation Metrics}

To summarize results, the metric of accuracy, precision, recall, and F1 are calculated for each artist.

The macro F1 is used to evaluate the performance of the multi-class classification problem. Besides, the macro of accuracy, precision, and recall are provided. The mathematical definition is shown in Equation (\ref{macro eq: acc f1 recall pre}).

\begin{equation}
\begin{aligned}
	\label{macro eq: acc f1 recall pre}
	Macro(m)=\frac{\sum_{n=1}^Nm_n}{N}
\end{aligned}
\end{equation}

\noindent where $m$ can be accuracy, precision, recall, and F1, $n$ represents the $n^{th}$ artist, and $N$ is the total number of the artists.

\begin{figure}[!htp]
	\centering
	\includegraphics[width=0.47\textwidth]{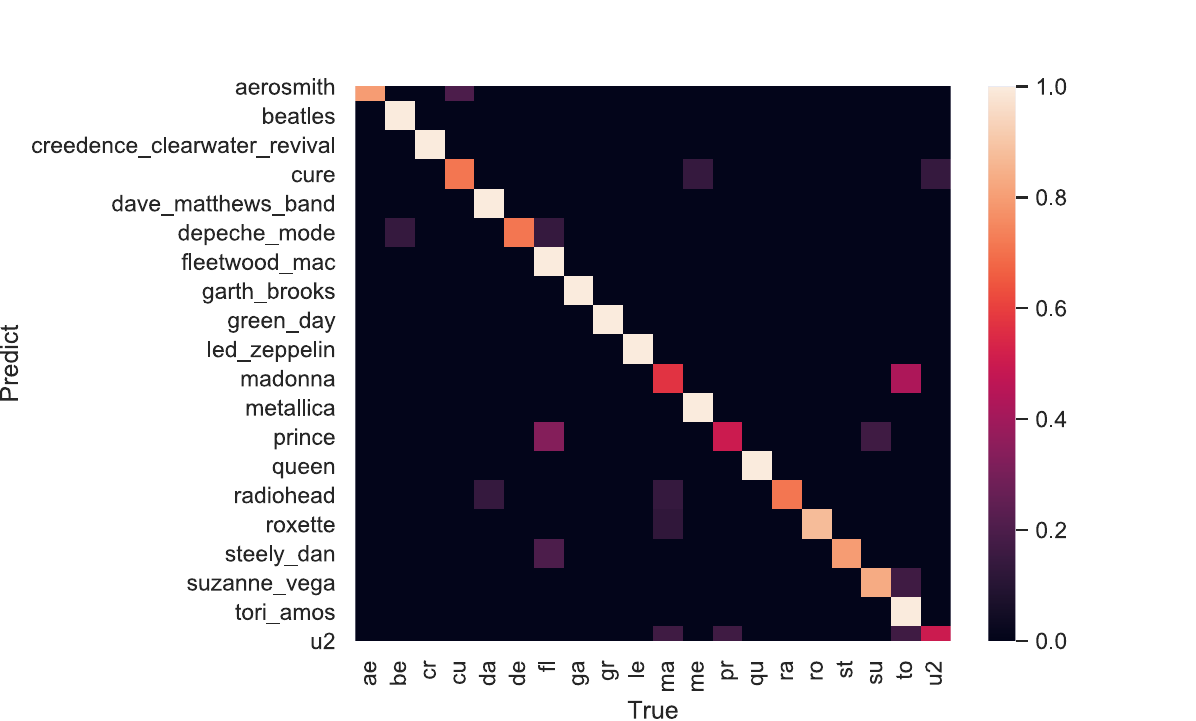}
	\caption{The confusion matrix of the song level evaluation on Artist20.}
	\label{fig:cm song}
\end{figure}

\subsection{Experimental Setup}

\begin{table*}[htbp]
	\centering
	\caption{The performance with different segment size as input on Artist20}
	\label{tab.differ segment size}
	\setlength{\tabcolsep}{3pt}
	\begin{tabular}{ccccccccc}
		\toprule

		                                                 & \multicolumn{4} {c}{\bfseries segment} & \multicolumn{4} {c}{\bfseries song}                                                                                                                            \\
		\cline{2-9}
		\multicolumn{1} {c} {\bfseries segment size (s)} & {\bfseries Accuracy}                   & {\bfseries Precision}               & {\bfseries Recall} & {\bfseries F1} & {\bfseries Accuracy} & {\bfseries Precision} & {\bfseries Recall} & {\bfseries F1} \\
		\midrule
		0.5 & 0.387 & 0.417 & 0.385 & 0.392 & 0.756 & 0.815 & 0.756 & 0.757 \\
		1.0 & \textbf{0.549} & \textbf{0.566} & \textbf{0.546} & \textbf{0.549} & \textbf{0.854} & \textbf{0.881} & \textbf{0.851} & \textbf{0.854} \\
		1.5                                              & 0.423                                  & 0.432                               & 0.431              & 0.424          & 0.632                & 0.634                 & 0.651              & 0.612          \\
		2.0                                              & 0.418                                  & 0.421                               & 0.488              & 0.428          & 0.595                & 0.591                 & 0.732              & 0.616          \\
	
		\bottomrule
	\end{tabular}

\end{table*}

\begin{table*}[htbp]
	\centering
	\caption{Comparisons with baseline models on Artist20 and Singer107}
	\label{tab.comparison}
	\setlength{\tabcolsep}{3pt}
	\begin{tabular}{ccccccccc}
		\toprule

	& \multicolumn{4} {c}{\bfseries Artist20} & \multicolumn{4} {c}{\bfseries Singer107}\\
		\cline{2-9}
		\multicolumn{1} {c} {\bfseries Method} & {\bfseries Accuracy} & {\bfseries Precision} & {\bfseries Recall} & {\bfseries F1} & {\bfseries Accuracy} & {\bfseries Precision} & {\bfseries Recall} & {\bfseries F1} \\
		\midrule
		i-vectors & 0.821 & 0.805 & 0.819 & 0.812 & 0.701 & 0.723 & 0.704 & 0.714 \\
		x-vectors & 0.832 & 0.835 & 0.831 & 0.829 & - & - & - & - \\
		MDCNN-SID & \textbf{0.854} & \textbf{0.881} & \textbf{0.851} & \textbf{0.854} & \textbf{0.796} & \textbf{0.804} & \textbf{0.823} & \textbf{0.816} \\
		\bottomrule
	\end{tabular}

\end{table*}

In this section, we will discuss the configuration of our experiments. The input size is an important factor for the performance of the proposed method. We have varied the input size from 0.5 seconds to 2 seconds with a step of 0.5 seconds. Due to the limitation of computing resources, the inputs of more than 2 seconds were not evaluated. For each input size, we evaluated our method on the validation set of Artist20.

As shown in Table \ref{tab.differ segment size}, the model reaches the best performance when the input size is settled as 1 second. Therefore, we choose the 1s as the input size in our experiments.

In the training stage, batch size and learning rate are set into 32 and 0.001, respectively. The Adam optimizer is adopted because it has shown strong performance in convolution-based classification tasks with limited hyper-parameter \cite{2019CRNN}. The early stopping mechanism is added with the patience of 10 to avoid over-fitting. The weights of the best model can be saved according to accuracy calculated on the validation dataset during the training.

\subsection{Experimental Results}

In this section, we compare our method with baseline methods, i.e., i-vectors \cite{2015ivector} and x-vectors \cite{snyder2018x}. Both two methods are used in speaker recognition tasks and have shown good performance.

We re-implemented these two methods and evaluated them on the datasets mentioned above. The comparison results of three different methods are arranged in the table \ref{tab.comparison}, where the proposed method is named MDCNN-SID. As is illustrated in the table above, The method we proposed has outperformed the baseline methods on every metric. Specifically, The score of Accuracy, Precision, Recall, and F1 of our proposed model are all four percentage points higher than that of the baseline model, which shows that the performance of our model has been greatly improved.

It should be noted that we do not compare our proposed method with other recent novel methods because computing resources are limited. According to the experiment result, our proposed method may not work better than all previous works. but it doesn't mean the method we proposed is useless, On the contrary, even in the case of limited computing resources, our model has still shown strong performance. which indicates that our idea is feasible, and our model is expected to show its real power in the case of sufficient computing resources in the future.     

The number of singers in Artist20 is far less than that of artists in the real world, which certainly can not satisfy the practical applications' requirements. To further evaluate our method in the case close to the real scene, we constructed and expanded the number of artists in the self-built Singer107 dataset, in which the artist category was expanded from 20 to 107 for comparison experiments as shown in Table \ref{tab.comparison}.

The result indicates that with the increase of artist categories, the performance of both baseline and our proposed method begin to degrade, but as shown in the table, we proposed method can still maintain high performance from the experimental results on the dataset of 107 singers(F1 score:0.816). The baseline method's recognition accuracy decreases more significantly as the number of artists increases.

The confusion metrics calculated on a test set at the song level are displayed in Figure \ref{fig:cm song}. In the confusion matrix, the diagonal lines mean that the points generated are based on the predicted results and the ground truth. The confusion matrix's vertical axis is the actual artist label, and the horizontal axis is the predicted artist label. Therefore, there is an obvious fact that the clearer the diagonals in the confusion matrix, the better the classification effect.

We can also observe that the trained model at the song level significantly improves the performance compared with the segment level since the model is trained by the short-time segment as input, while a complete song contains more than 180 segments. Classification errors are more likely to occur in the case of segment-level prediction since some segments do not contain apparent features, especially at the beginning and the end of a song. Voting by all the segments in the corresponding song can reduce the impact of error-prone segments, and the results also show that the precision and recall of the song-level evaluation have been improved.

\subsection{Discussion}

The task of artist classification is still a challenging task in MIR, which remains many difficulties to solve. Firstly, the singing voice to be recognized is inevitably intertwined with the background accompaniment. Recently, some researchers focus on addressing the confounds of accompaniments to improve the recognition accuracy \cite{sharma2019importance, hsieh2020addressing}. Secondly, the songs performed by music bands usually contain multiple singing voices, which can be hard to separate, and it will undoubtedly increase the difficulty of identification. Thirdly, artist classification's critical setting is each artist has a unique style and characteristic, which the model can learn. Nevertheless, in a real-life scenario, an artist can be influenced by another, collaborate with other artists' tracks, change the music style dramatically, etc. Those problems are widespread in certain kinds of music such as pop and electronic, where vocals are provided by featuring artists and similar styles.

Besides, data imbalance is also a crucial problem for the data-driven-based artist classification method. The number of albums and corresponding songs of the 20 artists in the dataset is balanced. The number of artists online is enormous, and the number of their albums and songs is imbalanced, which will be a tremendous challenge for model training. With the emergence of new artists, the model needs to continuously add new training samples for continuous updates and iteration.

\section{Conclusion}

In this paper,  a neural network model based on a multi-scale dilated convolution network is proposed for the artist classification task, which takes the raw audio waveform in the time domain as the input. Moreover, MDCNN-SID can enlarge the receptive field of convolution and integrate multi-resolution extracted features. The experimental results on the dataset Artist20 and Singer107 show that Compared with the baseline methods, the proposed method performs better, which indicates that the proposed method processing in the time domain can learn compelling features to distinguish artists. The effective combination of frequency domain and time domain is of great significance to music information retrieval. In the future, we plan to combine the raw waveform and spectrum of the audio signal to construct a hybrid model for artist classification tasks jointly.
\section{Acknowledgement}
This paper is supported by the Key Research and Development Program of Guangdong Province under grant No.2021B0101400003. Corresponding author is Jianzong Wang from Ping An Technology (Shenzhen) Co., Ltd (jzwang@188.com).

\bibliographystyle{IEEEtran}
\bibliography{mybib}

\end{document}